\documentstyle[12pt]{article}
\title{\Large\bf ORBITAL EVOLUTION OF A MASSIVE BLACK HOLE PAIR
BY DYNAMICAL FRICTION}
\author{ A{\sc lberto} V{\sc ecchio}$^{\,1}$$^{,\,2}$,
M{\sc onica} C{\sc olpi}
$^{\,2}$ \&
A{\sc lexander} G. P{\sc olnarev}$^{\,3}$$^{,\,4}$ \\
\\
$^{1\,}$
Dipartimento di Fisica Nucleare e Teorica, Universit\`a di Pavia.\\
Via Bassi 6 - 27100 Pavia - Italy.\\
\\
$^{2\,}$ Dipartimento di Fisica, Universit\`a degli Studi Milano.\\
Via Celoria 16 - 20100 Milano - Italy.\\
\\
$^{3\,}$ Astronomy Unit, Queen Mary and Westfield College,\\
University of London, Mile End Road, London, UK\\
\\
$^{4\,}$ Astro Space Centre of Lebedev Physical Institute,\\
Moscow, Russia}
\date{\today}
\begin{document}
\newcommand{\nab}{\mbox{$\vec\nabla$}}
\newcommand{\gras}[1]{\mbox{\boldmath $#1$}}
\newcommand{\ds}{\displaystyle}
\newcommand{\be}{\begin{equation}}
\newcommand{\ee}{\end{equation}}
\newcommand{\ba}{\begin{eqnarray}}
\newcommand{\ea}{\end{eqnarray}}
\newcommand{\m}{\langle}
\newcommand{\M}{\rangle}
\renewcommand{\theequation}{\thesection.\arabic{equation}}
\maketitle

\newpage

\section*{}
\centerline{ABSTRACT}
\vspace{10mm}
We investigate the evolution of a massive {\it black hole pair\/} under the
action of dynamical friction by
a {\it uniform\/} background of light stars with
isotropic velocity distribution.
In our scenario, the primary black hole
$M_1$
sits, at rest, in the center of the
spherical star distribution and the secondary
less massive companion $M_2$  moves along bound orbits determined
by the background gravitational field. $M_2$
loses energy
and angular momentum by dynamical friction,
on a time scale
longer than the orbital period.
The uniform star core
has total mass $M_c$ and radius $r_c$, and the following  inequality
$M_c>M_1>M_2$ holds.

In this paper, we investigate
mostly analytically the
{\it secular\/} evolution
of the orbital parameters, and  find that angular momentum ($J$)
and energy ($E$) are lost so as to cause the increase of
the eccentricity $e$ with time, during the orbital decay of $M_2$.

In the region of the core where the motion of $M_2$ is determined by the
mean field generated of the uniform stellar distribution, $E$ and $J$ are
lost {\it exponentially} on a time scale $\sim \tau_{\rm {DF}}$
determined by the properties of the ambient stars.
The rise of $e$ establishes instead on  a {\it longer\/} time
$\sim \tau_{\rm {DF}} \,(r_c/r_{\rm {A}})^2$  increasing as
the apocenter distance $r_{\rm A}$ decreases.

With the progressive decay of the orbit, $M_2$ enters the region
$r<r_B\sim (M_1/M_c)^{1/3}r_c$,
where the gravitational field of the primary black hole
dominates, but the star background maintains still uniform
(to first approximation).
Inside
$r_B,$
a key parameter of the calculation is the ratio between the black hole
velocity $v$ and the stellar dispersion velocity $\sigma.$~
(i)~ If $v< \sigma,$
energy and angular momentum
are lost {\it exponentially} on a time scale $\tau_{\rm {DF}}.$
The growth of $e$ occurs instead on a time  $\tau^e$ {\it longer} than
$ \tau_{\rm {DF}}$ by a factor
$\sim [\sigma/v]^2 .$  Therefore, $e$ rises
weakly during orbital decay.
$\tau^e$ is also found to be a function of $e$ and {\it increases}
as $e\to 1.$~
(ii)~ In the opposite limit, i.e., when $v>\sigma$,
the evolution of $E$ and $ {J}$ is close
to a {\it power-law} and establishes on a time scale
$\sim [v/\sigma]^3
\,\tau_{\rm {DF}}.$
The eccentricity grows on a time $\tau^e$
comparable to this scale.
Along an evolutionary path, $e$ increases significantly:
This rise leads the pericenter distance to diminish
exponentially, in this limit.
$\tau^e$ is a function of $e$ and {\it decreases} as $e\to 1$.
This limit ($v>\sigma$) is attained close to the cusp radius
$r_{\rm {cusp}}\sim
(M_1/M_c)\,r_c$, i.e., the distance below which the stellar distribution is
affected
by the gravitational field of $M_1$. Below $r_{\rm {cusp}}$
 our description is invalid, and we terminate our analysis.

Energy losses by gravitational wave emission become comparable
to those by dynamical friction at a critical distance that depends
on the ratio $M_1/M_c$: Consistency with the model assumptions
implies $M_1\ll M_c$.
The braking index $n$ is calculated in this transition region:
A measurable deviation from the value
of 11/3 corresponding to pure gravitational
wave losses provides ideally an indirect way for probing
the ambient medium.

\newpage

\renewcommand{\theequation}{3.\arabic{equation}}
\setcounter{equation}{0}
\section*{}
\centerline{1. INTRODUCTION}
\vspace{5mm}

Galaxies in their cores may contain massive binary black holes.
This hypothesis follows mainly from two considerations.
(i) Most galaxies harbor a central black hole,
relic of an earlier active phase
(Dressler 1989; Bland-Hawthorn, Wilson \& Tully 1991).
This assertion is plausible
if the AGN phenomenon
is the result of short-lived
phases of activity that have involved most of the galaxies in the past
(Rees 1990).
(ii) Galaxies interact and merge.
Evidence comes from
the appearance of unusual morphological features
interpreted as clear signature
of a tidal  interaction between companion galaxies or of a recent merger
(see Barnes \& Hernquits 1992 for a complete
review).
If galaxies participate a hierarchical process of clustering,
large galaxies cannibalize  the small ones and in these systems
a black hole binary
with unequal masses may form (Begelman, Blandford \& Rees 1980,
BBR hereafter; Roos 1981; Gaskell 1985).  Recently,
a claim was put forward for the observational
evidence of a massive binary system in the nucleus of
1928+738 (Roos, Kaastra \& Hummel 1993).

It is believed that
the nuclei of two interacting galaxies
sink, during a merger, toward the center of the
common potential well by dynamical friction:
a gravitationally bound pair of black holes
may thus form, embedded in core stars
(BBR; Valtaoja, Valtonen \& Byrd 1989; Governato, Colpi \& Maraschi 1993).
The frictional deceleration
on the  pair itself progressively tightens the system
and the two black holes eventually bind.
When their distance  becomes comparable to the mean star
separation,
single interactions with stars
on a time less than the orbital period come into play and
energy is transferred to the light particle. This process however
may lead to a depletion of background
stars that are driven into loss cone orbits
via a sling-shot mechanism (BBR; Roos 1981, 1988).
The decay of the binary
may be relented or even inhibited
at separations where the energy loss by emission of gravitational waves
would be weak to drive the binary to coalescence within the Hubble time.
Mechanisms for replenishing the loss cone
are therefore invoked for triggering further evolution
(BBR; Ross 1988; Polnarev \& Rees 1994).

Recent numerical $N$-body simulations
on the dynamical evolution of black hole binaries in the violently relaxed
core of a merged galaxy
(Fukushige, Ebisuzaky \& Makino 1992a;
Mikkola \& Valtonen 1992;
Makino et al.
1993), have indicated
that the collective effect of dynamical friction leads to a``runaway"
increase of the eccentricity
of the binary, and to a steep drop
of the pericenter distance.
Losses by gravitational wave emission would thus intervene early in
the life-time of the binary that evolves rapidly
(within $10^9$ yrs) toward final coalescence.
Proposing an approximate extension of the Chandrasekhar formula
for dynamical friction in a non-uniform anisotropic medium, Polnarev \& Rees
(1994; PR hereafter) studied numerically and analytically the evolution
of a black hole pair in a typical galaxy.
They find that in the core regions where
sufficiently large density gradients develop the eccentricity
decreases and velocity anisotropies tend to stabilize this tendency.

The study of the black hole
binary evolution is important since a
correct determination of the life-time and
occurrence rate of coalescence (Fukushige, Ebisuzaky \& Makino 1992b)
is crucial for assessing the relevance of black hole binaries
as candidate sources for the detection of gravitational waves
(Thorne \& Braginsky 1976; Thorne 1992). These sources could indeed be
detected by
Doppler experiments by spacecraft tracking
(Estabrook \& Wahlquist 1975) and long interferometer
in space as LISA (Danzmann et al. 1993) and
SAGGITTARIUS (Hellings et al. 1993).

Moving from these considerations, we  here
re-explore
the orbital evolution of a black hole pair, using mainly an
analytical technique.
In our model, a massive black hole $M_1$ sits,  at rest, in the center
of a {\it uniform galactic core\/} with total mass $M_c$, and radius $r_c$
(Binney \& Tremaine 1987).
A comparatively low mass black hole $M_2$ moves along bound
orbits spiraling
inward under the action of dynamical friction by the field stars.
We consider the relevant case when $M_2<M_1<M_c$ and introduce the
dimensionless
parameter $\eta\equiv M_1/M_c<1.$

Along the course of dynamical evolution,
the frictional deceleration on $M_2$
can be considered as small perturbation relative to
the ``background" gravitational
field, and therefore causes only {\it secular\/}
changes in the orbital motion.
In our scenario,
$M_2$ enters the core region of radius $r_c$
along bound orbits as it experiences the mean
gravitational field generated by the uniform
stellar distribution (see PR for detail). $M_2$ eventually bind to the massive
central black hole, at a distance $r_B\sim \eta ^{1/3}\,r_c$. Below
this separation, $M_2$ is still surrounded
by the homogeneous sea of field stars.
The  influence
of the central black hole on the stellar distribution becomes important at a
{\it smaller} separation $r_{\rm {cusp}}\sim r_c \eta$.
Below
$r_{\rm {cusp}}$ steep stellar density gradients and
anisotropies in the velocity field appear, and our treatment
ceases to be valid.

In this paper
we determine
the evolution of the orbital parameters
of $M_2$ in the two relevant regions: (a) the ``outer" zone,
in the interval $r_B<r<r_c$, where the
mean gravitational field is generated by the core stars and (ii)
the ``inner" zone $r_{\rm {cusp}}<r<r_B$ where the gravitational
field
of the massive central black hole affects the motion of $M_2$.

The plan of the paper is as follows.
In $\S 2$  the expression for
dynamical friction is reviewed and
key parameters of the model are introduced.
In $\S 3$ we
outline the basic equations for the energy and angular momentum
losses.  We solve these equations using
Gauss' perturbation theory,
both in the ``outer" and ``inner" core regions.
Characteristic time scales are
derived as a function of the eccentricity.
A criterion for the growth of $e$ is also presented and
applied to our analysis.
In $\S  4$ we describe
possible evolutionary paths the binary  experiences, within the inner
region of the galactic  core.
In $\S 5$ we review
formulae on the emission of gravitational waves:
the coordinated effect of
dynamical friction  and emission of gravity waves
is then discussed. The deviation, due
to dynamical friction, of the braking index
$n$ from the canonical value
of 11/3 is calculated.
$\S 6$ contains our conclusions.

\renewcommand{\theequation}{2.\arabic{equation}}
\setcounter{equation}{0}
\section*{}
\centerline{2. THE MODEL}
\vspace{5mm}

The frictional deceleration  experienced by the secondary BH
in a homogeneous background of stars with
isotropic gaussian velocity field is
\be
\dot{\gras{v}}_{{\rm DF}} = -\chi\, g(x)\, \hat{\gras{v}}\,,
\label{dyfr}
\ee
where
$\hat{\gras{v}}$ is a unit
vector in the
direction of $\gras{v}$, the velocity of the BH relative to the
background.

In equation (\ref{dyfr}), the
BH velocity is expressed in unit of the one-dimensional dispersion
velocity ($\sqrt 2\,\,\sigma$) of the stars
\be
x\equiv\frac{v}{\sqrt{2}\sigma}\,
\ee
and the function $g(x)$ reads
\be
g(x)=\frac{1}{x^2}\,\left[{\rm {erf}}
(x)-\frac{2}{\sqrt{\pi}}x e^{-x^2}\right]\,
\ee
(Chandrasekhar 1943; Binney \& Tremaine 1987).


The coefficient $\chi$ contains key parameters of the background stars
\ba
\chi & = & 2\pi\,\ln \Lambda\frac{G^2\,m\,n_0\,
(M_2+m)}{\sigma^2} \nonumber\\
& \sim & 10^{-8}
\left ({\ln \,\Lambda\over 10}\right )
\left ({M_2\over 10^6 M_{\odot}}\right )
\left ({m\over 1\,M_{\odot}}\right )
\left ({300\,{\rm {km\,s^{-1}}}\over \sigma}\right )^2
\,\,{\rm cm\,s^{-2}}\,,
\label{dfcost}
\ea
where $m$ ($\ll M_2$) is the mass of a typical individual star,
$n_0$ ($\sim M_c/m\,r_c^3$)  the number density of
stars and $\ln \Lambda=\ln(b_{max}/b_{min}) $
the so called Coulomb logarithm (Binney \& Tremaine 1987).

The  dissipative force (eq. [2.1])
parallel to the velocity vector $\gras{v}$
clearly acts so as to decelerate the secondary BH
causing the progressive decay of the binary separation.

Equation (\ref{dyfr}) takes a simple asymptotic expression
in two relevant limits.
In the first limit, i.e., for $v\ll \sigma $  (i.e., $x\ll 1$),
\be
\dot{\gras{v}}_{{\rm DF}} = -\frac{4}{3\sqrt{\pi}}\chi\,(x-\frac{3}{5}x^3+
o(x^3))\,
\hat{\gras{v}}.
\ee
This regime may characterize the early phase of the BH evolution
when orbiting regions where field stars move with
dispersion velocity $\sigma$ higher than $v$ (BBR).

In the second limit, i.e, for $v\gg \sigma$ (i.e., for $x\gg 1$),
the deceleration takes the form
\be
\dot{\gras{v}}_{{\rm DF}} \simeq -\chi\frac{1}{x^2}\,\hat{\gras{v}}.
\label{dyfrg}
\ee
The transition between the two asymptotic regimes
occurs when $x\simeq 1$, corresponding  to a value of
the mean separation between the BHs
$r_1 \sim \eta \, r_c$.

Simulations have shown that equation (2.1) often provides an
accurate description of the deceleration experienced by a single object.
In our model for the binary evolution, the drag acts only on the
secondary black hole since the primary is at rest relative to the
medium. If the primary were to move
a correction to equation (2.1) would be of need
(Bekenstein \& Zamir 1990; see however Gould 1993).

A source of uncertainty in equation (2.1) is the evaluation
of the Coulomb logarithm, and in particular the estimate of
the maximum impact parameter $b_{max}$ whose value is predicted
from numerical simulations
to be in the range $N^{-1/3} \,r_c$ (Smith 1992) and $r_c$ (Farouki \&
Salpeter 1982): We have chosen as reference value $\ln \Lambda=10$
(obtained considering $b_{max}=r_c$; Binney \& Tremaine 1987).
The deviation related to  a different choice of $b_{max}$
(i.e., a fractional error of less that $30\%$) falls anyway within the
intrinsic spread
of $\ln \Lambda$ related to the typical parameters of galaxy models.

Equation (2.1)
is considered valid
down to the limiting radius
$r_{{\rm {cusp}}}\simeq \eta\,r_c.$
At this distance the influence of the primary BH
on the stellar distribution is important
(Frank \& Rees 1976; Binney \& Tremaine 1987) and different mathematical tools
are of need to follow the subsequent evolution
(PR; see Maoz 1993 for a new approach to DF).

\vspace{7mm}
\centerline{2.2 {\it The Secondary Black Hole Orbital Motion}}
\vspace{5mm}

(a) The `` outer" region: In absence of dissipative forces the secondary BH
undergoes a
{\it periodic}
motion.
In the region $r_B<r<r_c,$  this motion is determined by
the gravitational potential $U(r)$ generated by
the uniform and isotropic
distribution of stars. According to the Poisson equation
\be
U(r) = \frac{1}{2}\Omega^2r^2\,,
\label{harmp}
\ee
where $\Omega$ is related to the density of stars $\rho=m\,n_0$ through the
following relation
\be
\Omega^2 = \frac{4\pi}{3}G\rho\,.
\ee
The secondary BH undergoes a periodic motion along elliptic orbits:
In cartesian coordinates, with
origin in the center of the star distribution, the motion is described by
\be
x = r_{{\rm A}} \cos(\Omega t)\,,
\ee
\be
y = r_{{\rm P}} \sin(\Omega t)\,,
\ee
where
$r_{{\rm A}}$ and $r_{{\rm P}}$ are the apocenter and pericenter distances,
respectively.

We can express the two constants of motion, i.e.,
the total energy $E$ and angular momentum $J$
in the following form
\be
r_{{\rm A}}^2+r_{{\rm P}}^2 = 2\frac{E}{M_2\Omega^2}\,,
\ee
\be
r_{{\rm A}}\,r_{{\rm P}}=\frac{J}{M_2\Omega}\,,
\ee
and introduce the eccentricity
\be
e = \frac{r_{\rm {A}}-r_{\rm {P}}}{r_{\rm {A}}+r_{\rm {P}}}\,.
\ee
For the analysis of the motion of the secondary BH driven by DF it is
of importance to estimate the magnitude of the deceleration
${\dot {v}}_{\rm {DF}}$ and compare it with the acceleration
generated by the gravitational potential
${\dot {v}}_{\rm {H}}=\Omega^2\,r$. In  the limit
$x\ll 1$, we have
\ba
\dot{{v}}_{{\rm DF}}\over \dot{{v}}_{\rm {H}} & = &
\frac{4}{\sqrt{3}}G\,M_2\ln{\Lambda}\sqrt{G\rho}\,\sigma^{-3}\nonumber \\
& =& 5\times 10^{-3}\left(\frac{M_2}{10^6M_{\odot}}\right)
\left(\frac{M_c}{2\times 10^8\, M_{\odot}}\right)^{1/2}
\left(\frac{100\, {\rm {pc}}}{r_c}\right)^{3/2}
\left(\frac{300\, {\rm {km\,s^{-1}}} }{\sigma}\right)^3
\ea
The weakness of the frictional drag enable us to treat DF as
a {\it perturbing\/} force.

For the harmonic potential, we consider only the case $x\ll 1$.
{}From the virial theorem
and from equation (2.8) this condition is equivalent to
\be
\frac{r_{{\rm A}}}{r_c}< 1
\ee
which is obviously satisfied. The opposite limit ($x\gg 1$), equivalent to
$r_{{\rm P}}/r_c> 1$,
corresponds to a motion confined in regions that are external
to the galaxy core. This limit is
therefore inconsistent.

(b) The ``inner" region: The secondary BH becomes eventually bound
to the primary BH. This occurs  at a separation
\be
r_B\sim
\eta^{1/3}\, r_c\sim 37
\left ({M_1\over 10^8M_{\odot} }{2\times 10^9\, M_{\odot}\over M_c}
\right )^{1/3} \left({r_c\over 100 \,{\rm pc}}\right)\,\,{\rm pc}
\ee
for typical parameters of a galactic core (BBR).
At radii $r< r_{{\rm B}},$ the potential of the
central BH dominates over the mean potential
generated by the homogeneous see of stars. At $r_B$ the
BH binary forms
and the secondary BH moves along
Keplerian orbits.
In the region $r_{\rm {cusp}}<r<r_B$, the binary is
surrounded  by ambient stars having
a uniform density and dispersion velocity $\sigma$ larger than $v$.
This is the reason why only below the cusp radius
\be
r_{\rm {cusp}}\sim
\eta\, r_c\sim 1
\left ({M_1\over 10^8M_{\odot} }{2\times 10^9\, M_{\odot}\over M_c}
\right ) \left({r_c\over 100 \,\rm {pc}}\right)\,\,\rm {pc}
\ee
the stellar distribution is affected by the central BH.

The total newtonian energy of the binary
\be
E=-{GM\mu\over 2a}
\ee
and angular momentum $\gras{J}$ of magnitude
\be
J^2=GM\mu^2a(1-e^2)\,
\ee
are constants of motion
(in eq. [2.18] and [2.19] $M=M_1+M_2\sim M_1$ and $\mu\sim M_2$
is the reduced mass). Given $E$ and $J$
the semimajor axis $a$
and the eccentricity $e$ are uniquely determined.

Analogously to the previous case, we estimate the
magnitude of the frictional drag ${\dot v}_{{\rm DF}}$
relative to the gravitational acceleration,
$\dot{{v}}_{{\rm K}}=GM/a^2$. For $x\ll 1$ we have
\ba
{\dot{{v}}_{{\rm DF}}\over \dot{{v}}_{\rm {K}}} & = &
{4\over 3\pi^{1/2}} {\chi\over \sigma} {a^{3/2}\over
(GM)^{1/2} }
\nonumber\\
& \sim & 2.2\times 10^{-3}
\,\,\,\left ( {a\over 10\, \rm {pc}}\right )^{3/2}
\,\,\left ({300\, {\rm {km\,s^{-1}}}\over \sigma} \right )^3\,\,
\left ({10^8\, M_{\odot}\over M}\right )^2\,
\ea
and for $x\gg 1$
\ba
{\dot{{v}}_{{\rm DF}}\over \dot{{v}}_{{\rm K}}}
& = &
2\chi \sigma^2 {a^3\over (GM)^2}\nonumber\\
& \sim & 2.3\times 10^{-8}\,\,\,
\left ({a\over 0.1\, \rm {pc} }\right )^3
\,\,\left ({ 10^8 M_{\odot} \over M}\right )^2\,.
\ea
Even in this case DF is weak compared to gravity.
Therefore, DF can be considered as a perturbing force that controls
only the evolution
of the  BH on a time scale longer than a orbital period.
In both potentials, the motion of the secondary BH is almost periodic.
On the secular scale fixed by DF, the mean motion
varies,  the system evolving
toward states that are progressively more bound.

\renewcommand{\theequation}{3.\arabic{equation}}
\setcounter{equation}{0}
\section*{}
\centerline{3. THE LONG TERM EVOLUTION OF THE SECONDARY BH}
\vspace{5mm}
\centerline{3.1 {\it Long Term Variation of $E$ and $J$}}
\vspace{5mm}

In this section we derive the long term evolution of
$E$ and $J.$
In presence of DF, energy
is not conserved and it is dissipated
at a rate
\ba
\dot{E} & = & \mu \,\dot{\gras{v}}_{{\rm DF}} \cdot \gras{v}
\nonumber\\
& = & -\mu \,\chi\, g(x)\, v\,.
\label{enper}
\ea
Similarly, the equation for the angular momentum loss is
\ba
{\dot {\gras J}} & = &
\mu ( {\gras {r}}\times \dot {\gras{v}}_{{\rm DF}} )
\nonumber\\
& = & -\chi g(x) {{\gras {J}}\over v}\,.
\label{maper}
\ea
\noindent
$\dot E$ and $\dot J$ represent the  instantaneous values
of the time derivatives for $E$ and $J$. Since DF is a weak perturbing
force, it is meaningful to calculate $\m{\dot E}\M$
and $\m{\dot J}\M $, i.e., the average
on the orbital period.
This mean is defined for any orbital parameter $S$ (such as $E,J$,
$r_{{\rm P}}$, $r_{{\rm A}}$, $a$ and $e$) as
\be
\m\dot{S}\M \equiv \frac{1}{P}\int_0^P\!\dot{S} dt\,,
\ee
where
$P$ is the orbital period.
Below, the results are presented
separately in the two asymptotic limits of equation (2.1) and for the
two expressions of the gravitational potential acting on the secondary BH:
the calculations are fully analytical.

\vspace{5mm}
\centerline { (a) {\it The Harmonic Potential} }
\vspace{5mm}

When the gravitational potential of the star distribution is dominant
($r_B<r<r_c$),
the losses can be evaluated for $x\ll 1$ retaining only the
first term in the expansion of $g(x).$
This yields:
\be
\m \dot{E}\M = -\frac{4\chi}{3\sqrt{2\pi}\sigma} E
\ee
\be
\m \dot{J}\M = -\frac{4\chi}{3\sqrt{2\pi}\sigma} J\,.
\ee
It is relevant to notice that in this limit
the decay is {\it exponential}
as DF is proportional to $\m v \M$ in this limit (see eqs. [2.5], [3.1]
and [3.2]). These losses are also to lowest order
independent of the eccentricity $e.$

As indicated by equations (3.4) and (3.5),
the magnitude of the decay time is primarily
determined by the values of $\chi$ and $\sigma$.
We are therefore led to introduce a characteristic time scale
for DF
\ba
\tau_{\rm {DF}} & \equiv & \frac{\sqrt{2}\sigma}{\chi}\nonumber\\
& \simeq &
10^7 \left(\frac{\sigma}{300\,
\rm {km \,s}^{-1}}\right)^3
\left(\frac{10^6 M_{\odot}}{M_2}\right)
\left ({2\times 10^3 {\rm \,\,{pc^{-3}}}\over n_o}\right ){\rm yrs} \,.
\ea
In terms of $\tau_{\rm {DF}}$
the decay time scales for $E$ and $J$ are equal and
read:
\be
\tau_{H}^E = \tau_{H}^J = \frac{3\,\pi^{1/2}}{4} \tau_{\rm {DF}}\,.
\ee

\vspace{5mm}
\centerline { (b) {\it The Keplerian Potential} }
\vspace{5mm}
Below $r_B$ the gravitational potential of the massive
primary BH dominates and the motion of the secondary is nearly
Keplerian.

\vspace{7mm}
\centerline { (i) {\it The limit $x\ll 1$} }
\vspace{5mm}

We consider the case of
$x\ll 1$ and, for simplicity, retain again  the
first term in the expansion of $g(x).$
In this case the losses of energy and angular momentum read
\be
\m\dot{E}\M = -\frac{8\chi}{3({2\pi})^{1/2}\sigma} E
\ee
\be
\m\dot{J}\M = -\frac{4\chi}{3 ({2\pi})^{1/2}\sigma} J
\ee
(the details are presented in Appendix A and Appendix B. In the calculations
$x$ is a function of the orbital phase and the asymptotic limit
is meant to be valid over the complete orbit).
It is important to notice that even in this case the losses are
{\it exponential} and independent of the eccentricity.
According to the previous
equations the corresponding time scales are
\ba
\tau_{<}^E
& = & \frac{1}{2} \tau_{<}^J\nonumber\\
& = & \frac{3\,\pi^{1/2}}{8} \tau_{\rm {DF}}\,.
\ea

\vspace{5mm}
\centerline { (ii) {\it The limit $x\gg 1$} }
\vspace{5mm}

With the progressive loss of energy to the ambient stars,
the orbital velocity of the secondary BH may
increase above $\sigma$ (see $\S 4$ for a discussion).
For $x\gg 1$,
the evolution equations, after considerable manipulations,
can be cast into the following form
\ba
\m\dot{E}\M & = & -\frac{\chi\sigma^2\mu^{3/2}}{2^{1/2}\pi}\,E^{-1/2}\,\,
{\cal K}(e)\nonumber\\
& = & -\frac{\chi\sigma^2\mu}{\pi}
\left( \frac{a}{GM}\right )^{1/2}\,\,{\cal K}(e)\,,
\ea
\be
\m\dot{J}\M  = -\frac{\chi\sigma^2\mu}{\pi}\frac{a^2}{GM}\,\,{\cal Z}(e)\,,
\ee
where in ${\cal K}(e)$, and ${\cal Z}(e)$
we confine the dependence on
$e$ (the details are presented in Appendix B).
Equations (3.11) and (3.12) show that the time evolution of
$E$ and $J$ is not exponential in this limit but
follows approximately a {\it power--law}.

These equations contain the relevant
time scales $\tau_{>}^E$ and $\tau_{>}^J$ that
depend, in this limit,  on  $e$ and
on a new quantity
\be
x_{\rm {cir}} \equiv \left(\frac{G\,M}{2\sigma^2 a}\right)^{1/2}\,,
\ee
representing the velocity the BH would have
as if it were to move along a circular
orbit; this velocity is expressed in units of the dispersion velocity of the
background stars. In $x_{\rm {cir}}$ we mainly retain the dependence
of the time scale  on the energy $E$ (see eq. [2.18]).

The resulting  time scales are
\ba
\tau_{>}^E & \sim &
 x_{\rm {cir}}^3\,\tau_{\rm {DF}}\,\frac{1}{{\cal {K}}(e)}\nonumber\\
& \sim & 5\times 10^8 \left(\frac{x_{\rm {cir}}}{3}\right)^3
\frac{1}{{\cal {K}}(e)}\,{\rm yrs}\,,
\ea
\ba
\tau_{>}^J & \sim & x_{\rm {cir}}^3\,\tau_{\rm {DF}}\,
\frac{(1-e^2)^{1/2}}{{\cal {Z}}(e)}\nonumber\\
& \sim & 5\times 10^8 \left(\frac{x_{\rm {cir}}}{3}\right)^3
\frac{(1-e^2)^{1/2}}{{\cal {Z}}(e)}\, {\rm yrs}\,.
\ea
It is of interest to notice that, in this regime, the time scales
of orbital decay $\tau^E_>$ and $\tau^J_>$
are {\it longer} than
the ``reference" time scale $\tau_{\rm {DF}}$ (i.e., the scale for
$x\ll 1$), and the dilution factor is of $\sim x^3,$
as DF on $\propto v^{-2}$ (equations [2.6]
and [3.1]).
Considering further their dependence on $e$ through $\cal K$ and $\cal Z,$
both time scales are found to
become progressively {\it shorter} with increasing eccentricity.
In Figure 1, we show $\tau_{>}^E$ and $\tau_{>}^J$  versus $e$
and note the decrease of the time scales as $e\to 1:$
As an example for $J$,
$\,\tau_<^J \,(e=0.9)\sim 0.1\,\tau_<^J (e=0).$

The release of energy by the decaying binary is
a source of heating for the background. For our
choice of parameters, however, the relative enhancement of the
dispersion velocity is negligible, being $\Delta\,\sigma/\sigma\sim
G\,M_1\,M_2/\eta\,r_c/N\,m\,\sigma^2\sim M_2/M_c\ll 1$.

\vspace{5mm}
\centerline{3.2 {\it The Criterion for the Growth of the Eccentricity}}
\vspace{5mm}

In the decay of the orbit,  energy
and angular momentum losses
to the field stars may lead either to the {\it growth}
or the
{\it decrease} in the eccentricity of the system, the effect having
important consequences on the life-time of the binary.
We here formulate two simple and general
criterions
for the growth of $e$, the first applies to the motion in the harmonic
potential and the second to the
motion in the Keplerian potential.

\vspace{5mm}
\centerline{(a) {\it The Harmonic Potential}}
\vspace{5mm}

{}From equation (2.13) we can express the eccentricity in the following form:
\ba
e^2 & = & \left(\frac{r_{\rm A}-r_{\rm P}}{r_{\rm A}+r_{\rm P}
}\right)^2\nonumber\\
& = &
\left ( {E\over \Omega}-J\right ) \left ({E\over \Omega}+J\right )^{-1}
\,.\ea
The time derivative of the eccentricity therefore reads
\be
\frac{de^2}{dt} = \frac{2}{\Omega}
(\dot{E}J-E\dot{J})
\left[\frac{E}{\Omega} + J\right]^{-2}\,.
\ee
It is obvious that the sign of $\dot e$ depends on the competition
between the energy loss and the angular momentum loss, i.e., on the sign of
$\dot{E}J-E\dot{J}.$ We are therefore
led to introduce a general criterion for
the growth of $e$ using equation (3.17). If the condition
\be
\m \dot{E}J-E\dot{J}\M  > 0
\ee
is satisfied along the BH orbital evolution, the eccentricity will rise.

Substituting into equation (3.18) the expressions of the energy
and angular momentum losses (eqs.[3.4] and [3.5])
it is straightforward to verify
that $\m \dot {e} \M=0$. The eccentricity remains thus frozen to the value
set at the outset of the BH evolution.
This holds if only the lowest order in $x$ is
retained in equation (2.5). If we include the next leading term in
the expansion of $g(x)$ we  obtain
\be
\m \dot{E}J-E\dot{J} \M = \frac{\chi}{20\sqrt{2\pi}\sigma^3}\,J\Omega^2\,
(r_{{\rm A}}^2-r_{{\rm P}}^2)^2\,.
\ee
The r.h.s. of equation (3.19) is definite positive.
We thus find that the eccentricity
increases during the first stage of the
orbital motion of the secondary BH.

We can also calculate the rate of increase of $e$
\be
\m \dot e \M = \frac{1}{5}\frac{\chi}{\sqrt{2\pi}\sigma^3}\,\Omega^2
(r_{{\rm A}}^2-r_{{\rm P}}^2)
\ee
and the associated time scale
\ba
\tau_{{\rm H}} & = & 5\sqrt{2\pi} \tau_{{\rm DF}}
\frac{r_c^2}{(r_{{\rm A}}+r_{{\rm P}})^2}\nonumber\\
& \sim  & 6\times 10^7 \left(\frac{r_c}{100 {\rm pc}}\right)^2
\left(\frac{70 {\rm pc}}{r}\right)^2\,{\rm yrs}.
\ea
Note that $\tau_{{\rm H}}$
exceeds $\tau_{\rm {DF}}$ by a factor $\sim (r_c/r_{\rm {A}})^2:$
This indicates that the rise of $e$ is not severe in this evolutionary phase
as
it occurs on a time scale {\it longer} than
the time  for energy and angular momentum losses.

\vspace{5mm}
\centerline{(b) {\it The Keplerian Potential}}
\vspace{5mm}

In the region $r_{\rm {cusp}}<r<r_B$,
the orbital motion of the secondary BH is described by the
equations (2.18) and (2.19). They relate $e$ to $J$ and $E$ to give
\be
J^2=-\mu^3\frac{G^2M^2}{2E}(1-e^2)\,.
\label{momang1}
\ee
We can thus derive the
expression for the time derivative of the eccentricity
\be
\dot{e}  =  -\frac{1-e^2}{e}\left(\frac{\dot{J}}{J}+\frac{1}{2}
\frac{\dot{E}}{E}\right)\,.
\label{eccdot}
\ee

Similar to the previous case,
we can introduce a general criterion
for the growth of  the $e.$
Using equation (\ref{eccdot})
the condition is
\be
\frac { \m\dot{E}\M}{\m\dot{J}\M}<-\frac{2E}{J}\,.
\label{einst}
\ee
Below, we apply this criterion of growth using the
results of $\S\,\,3.1.$

\vspace{5mm}
\centerline {(i)~{\it The limit $x\ll 1$}}
\vspace{5mm}

Equations (3.8) and (3.9) indicate that
the eccentricity is constant along the motion
(i.e., $\m\dot {e}\M=0$) since $\m\dot {E}\M/\m\dot {J}\M=-2E/J$ sharply.
This  result holds if only the lowest order in $x$ is retained
in equation (2.5).
If we include the next leading term in the expansion of $g(x)$,
the long term changes of the energy
and angular momentum are found to be characterized by a ratio
\ba
\frac{\langle \dot{E} \rangle}{\langle \dot{J} \rangle}  \simeq
-\frac{2E}{J} &  & \times \left[ 1-\frac{3}{20\pi\sigma^2}\frac{GM}{a}
(1-e^2)^{-1/2}I_2 \right]\nonumber\\
& & \times \left[1-\frac{3}{20\pi\sigma^2}\frac{GM}{a}
(1-e^2)^{1/2}I_1\right]^{-1}\,,
\ea
that depends now explicitly on $e$
(the integrals $I_1$ and $I_2$ are given in Appendix B).
{}From equation (3.18),
the criterion for the ``growth'' of $e$ is fulfilled
if
\be
1  <  \frac{1}{1-e^2}\frac{I_2(e)}{I_1(e)}
= \frac{4-3(1-e^2)^{1/2}}{(1-e^2)^{1/2}}\,.
\ee
This inequality holds for $e^2>0$, and  equation (3.26) is thus
satisfied $\forall e\in [0,1[.$
In the limit $x\ll 1$ we thus find that DF on the secondary BH acts so as to
increase the eccentricity, during orbit decay.
The time scale for the growth of $e$ can be estimated
considering the evolution equation for the eccentricity.

Following Bertotti and Farinella (1990), we compute the secular change of $e$
(see Appendix C). After some calculations, we are able to
find a simple analytical
expression
\be
\m\dot{e}\M = \frac{8\chi}{5(2\pi)^{1/2}\sigma^3}\,\frac{G\,M}{a}\,
\frac{(1-e^2)^{1/2}}{e}\,\left[1-(1-e^2)^{1/2}\right]\,.
\ee
Note that $\m\dot{e}\M$ is positive definite for any value of $e,$
and the
characteristic time of growth for  $e$
is
\ba
\tau_{<}^e & \sim & \frac{e}{\m\dot{e}\M} = \frac{5\pi^{1/2}}{8}\,
\, {\tau_{{\rm DF}}\over x^2_{\rm {cir}}}
\,\,
\frac{e^2}{2(1-e^2)^{1/2}\left[1-(1-e^2)^{1/2}\right]}\, {\rm yrs}\nonumber\\
& \sim & 7\times 10^8 \left(\frac{0.3}{x_{\rm {cir}}}\right)^2
\,\,\frac{e^2}{2(1-e^2)^{1/2}\left[1-(1-e^2)^{1/2}\right]} \,{\rm yrs} \,.
\ea
The behavior of $\tau_{<}^e$ is reported in Figure 2.
$\tau_<^e$ is found to {\it increase} as $e\to 1$
but the deviation is significant only for extreme values of $e$,
very close to unity.
This time scale is {\it longer} than the
scale on which
energy and angular momentum are lost (eq. [3.10]), and the  dilution factor
is $\sim 1/x_{\rm {cir}}^2$. Hence, only a weak rise of
$e$ is expected to
take place, in the early phase of binary decay.

\vspace{5mm}
\centerline {(ii) {\it The limit $x\gg 1$}}
\vspace{5mm}

In this limit, the ratio $\m \dot {E} \M / \m\dot {J} \M$ reads
\ba
\frac{\langle \dot{E} \rangle}{\langle \dot{J} \rangle} & = &
\left(\frac{GM}{a^3}\right)^{1/2}\frac{{\cal K}(e)}{{\cal Z}(e)}\nonumber\\
& = & -\frac{2E}{J}\left[\frac{(1-e^2)^{1/2}{\cal K}(e)}{{\cal Z}(e)}\right]\,.
\ea
Considering the dependence of ${\cal K}(e)$ and
${\cal Z}(e)$ (drown in Figure B1 and B2 of
Appendix B) we find that
criterion (3.24) is satisfied
$\forall e$. As a result, $e$ {\it increases}
since angular momentum is lost by DF more effectively than energy, as
for the opposite limit.

In particular, at small eccentricities ($e\ll 1$) the ratio
${\langle \dot{E} \rangle}/{\langle \dot{J} \rangle}$
takes a simple form
that we derive in Appendix B:
\ba
\frac{\langle \dot{E} \rangle}{\langle \dot{J} \rangle} & \simeq &
 -\frac{2E}{J}(1-3e^2)\nonumber\\
& < & -\frac{2E}{J}.
\ea

{}From equations (C.1) and (2.6), the
long term change of $e$ is governed by the following equation (for
the details see Appendix C)
\be
\m\dot{e}\M = \frac{2\chi\sigma^2}{\pi}\left(\frac{a}{GM}\right)^{3/2}
{\cal {W}}(e)\,,
\ee
where ${\cal {W}}(e)$ is a positive function given in Appendix C.

${\cal W}(e)$ takes a very simple form in the case $e\ll 1$
\be
{\cal {W}}(e) = \frac{3}{2} e\left[1+\frac{3}{8}e^2+o(e^2)\right].
\ee
The associated  time scale is
\ba
\tau_{>}^e & \sim & x_{\rm {cir}}^3\,\tau_{{\rm DF}}
\frac{e}{{\cal {W}}(e)}\nonumber\\
& \sim & 4 \times 10^8 \left(\frac{x_{\rm {cir}}}{3}\right)^3
\frac{e}{{\cal {W}}(e)}\,{\rm {yrs}}\,.
\ea
The behaviour of $\tau_{>}^e$
is shown in Figure 3.
At small eccentricities the time scale is
\be
\tau_{>}^e\sim \frac{1}{3}x_{\rm {cir}}^3\,
\tau_{{\rm DF}}
\left[1-\frac{3}{8}e^2+o(e^2)\right].
\ee
Contrary to the case for
$x\ll 1,$
the evolution time is found to decrease with increasing $e:$
Thus, eccentric orbits become more eccentric on a progressively shorter
time scale: We may refer to this effect  as ``runaway''
increase of $e$.

\vspace{5mm}
\centerline {(iii){\it  $\,\,\,\,x\sim 1$ } }
\vspace{5mm}

We have studied so far
the long term evolution
of $E,J$ and $e$
in the two asymptotic limits
of $x$. To complete our analysis
we explore numerically the evolution
equations (3.1) and (3.2) using the complete expression of $g(x).$

The long term evolution of $E$ and $J$ when $\m x\M \simeq 1$ is
found to be intermediate the two bracketing limits.
Accordingly, we find that the eccentricity
{\it increases\/}  during orbit decay since angular momentum is always
lost
more effectively than energy.
Whereas in the limit $\m x\M \gg 1$ the losses at pericenter are weaker
than at apocenter (eq.[2.6]) the opposite
holds when $\m x \M \ll 1$ (eq.[2.5]). Nonetheless
also in the last case, $\m\dot {e}\M$ maintains the same sign.
Pervious arguments based on comparisons between losses
at different orbital phases were used to derive conclusive
statements
for the growth of $e$;
the above finding thus shows the weakness
of that approach.

We summarize the results of numerical integrations
in Figure 4 , where
$\m{\dot E}\M/\m{\dot J}\M$  in units of $-2E/J$ is shown as a function of $e,$
for selected values of $x_{{\rm cir}}$.
It is remarkable that regardless the specific value of $x$,
this ratio maintains below unity over the complete
domain [0,1[, and becomes smaller with either
increasing eccentricity (close to unity) or increasing $x$.
This is in accordance with the behavior of
$\tau^e$ (eqs. [3.28], and [3.33]).

\renewcommand{\theequation}{4.\arabic{equation}}
\setcounter{equation}{0}
\section*{}
\centerline{4. POSSIBLE EVOLUTIONARY PATHS FOR THE BH BINARY}
\vspace{10mm}

In $\S 3$ we have shown that the BH eccentricity
increases slowly
in the region $r_B<r<r_c$ where the gravitational field
of the core dominates.
As a consequence,  about  $r_B,$ $e$
has a value close
to the one  acquired about $r_c$.
The orbital parameters thus keep memory of the previous BH evolution
in regions $r>r_c$.
The action of DF
at radii $r>r_c$ where the density of the stellar distribution
decreases nearly as $r^{-2}$
have been investigated numerically in PR.
In this work it is shown that
the BH reaches $r_c$ with a small eccentricity: It follows that
the
BH trajectory  is ``prepared"  to have $e\ll 1$ at $r_B.$

Starting from these considerations, in this section we  explore
the long term evolution of the secondary BH and in particular the evolution
 of the
eccentricity $e(E,J)$ in the ``inner" core region  where the ``instability"
that drives a bound orbit to nearly radial infall
($e\to 1$) may set in.

Consistency with the model assumptions of a uniform
star background requires the distance of closest approach
not to
drop below $r_{\rm {cusp}}$, around $M_1$ (i.e., around
the focus of the ellipse).
In an equivalent way, the apocenter should not exceed $r_B$
for the motion to be consistent with
the Keplerian potential.  These constraints lead to the following
inequalities
(for $a$ and $e$)
\be
a\,(1+e)<\eta^{1/3}\,r_c
\ee

\be {a\,(1-e)(1+e)\over e}>\eta \, r_c\,.
\ee

For sufficiently high values $e$ and binding energy $-E$, the BH can probe
the region within the cusp where density gradients
are present and where $\sigma\simeq v;$  this region is not accessible
to our model.

\vspace{5mm}
\centerline{4.1 {\it Selected Paths}}
\vspace{10mm}

We begin the  numerical integration of the evolution
equations (3.1-3.2) at  a distance
$a_i= 20\,{\rm pc}$ (for which $\m x_i \M \simeq 0.3$) and choose a value
of $\eta\simeq 0.05$ (following BBR).
Along the course of the BH evolution $\sigma$ is kept constant (see $\S 3.1$).

The eccentricity is
treated as a free parameter; small initial
values are preferred being consistent
with the previous BH history (PR).
In the evolution equations, the
drag by the ambient stars is  computed using the complete
expression of $g(x)$ (eq.[2.3]). The means (eq. [3.3]) are evaluated
numerically and a test on the accuracy of the calculation is performed
using the analytical results obtained in the asymptotic limits.

Figure 5 shows possible paths for $e$ as a
function of the semimajor axis $a$ which can be considered in place of the
time coordinate.
It appears that,
independent of the initial value of the eccentricity
$e_i$, the growth of
$e$ with decreasing $a$ is not severe as long as the BH velocity
does not exceed the dispersion velocity of the ambient stars
(i.e, until $\m x \M < 1$).
At separations where $\m x \M \sim 1,$
the value of $e$
still depends on $e_i$ and the binary keeps memory
of the initial condition.
The ``runaway growth'' of $e$ occurs only
as the mean value of $\m x \M$ increases just above unity.
In Figure 5, the function $e(a)$ is tracked also for $a<r_{\rm {cusp}}$
(dashed line). In this case
the pericenter distance $r_{\rm {P}}$
falls within the cusp radius, and the above inequalities
(eqs. [4.1] and [4.2]) are violated.
The path below $r_{\rm {cusp}}$ is shown just to indicate how rapid would
be the growth of $e$
if the BH were to move in a uniform medium even below $r_{\rm {cusp}}$.

We now consider an evolutionary path $a(t)$ having $a_i=20$~pc, and
a small $e_i=0.03$, to follow the evolution
of the pericenter  $r_{\rm P}(t)\equiv a\,(1-e)$
and apocenter  $r_{\rm A}(t)\equiv a\,(1+e)$ distances with time.
According to the analysis of  $\S 3$
we find that in the ``inner" core region
where the BH velocity $\m v \M < \sigma$,
energy and angular momentum are lost to the background stars
exponentially, on a time scale $\tau _{{\rm DF}}\sim 10^7\,{\rm yrs}$.
In this phase, the semimajor axis decreases exponentially
but
$e$ does not vary appreciably
having a longer time scale of growth
($\propto \tau_{{\rm DF}}/\m x \M ^2$; cfr. eq. [3.28]):
Accordingly $r_{\rm {P}}(t)\sim r_{\rm {A}}(t)\sim a(t).$
The binary evolves rapidly through this
early phase, until
$\m v \M \sim \sigma$.  The dynamical evolution will proceed from
this moment onward on a different time scale
$\sim \m x\M ^3\,\tau_{{\rm DF}}
\sim 10^8\, {\rm yrs},$ since DF ($\propto v^{-2}$)
weakens as the binary hardens. The increase of the stellar density
(by a factor $\sim 2.5$ for a density profile $\sim r^{-7/4}$; see Frank \&
Rees 1976) would however partly compensate this delay in the BH evolution.

In summary, our idealized study shows that
we can distinguish two evolutionary phases for the BH ``binary": (1) an early
phase, lasting a few $10^7$ yrs, in which the inspiral
of the two BHs occurs along bound orbits of
slowly increasing eccentricity (the binary is `` nearly stable''
to the rise of $e$). (2)~ a subsequent phase, not described by our model,
in which the BH probes the cusp region at pericenter, and the
uniform core at apocenter. The non-local nature of DF prevents us
to give a correct estimate of the evolution of the eccentricity, during
the BH inspiral.
According to recent studies,
the eccentricity would remain nearly constant
in the region below the cusp where stars
can still be treated as a continuum medium (see PR).
Where collisions with single stars come into play,
$e$ can instead increase again, as indicated in
numerical experiments by Roos (1981) and Hills (1983).
Hence, the evolution of $e$ depends sensitively on the detail of the
interaction with stars in the innermost region of the galactic core,
before losses by gravitational waves drive the binary to
final plunge.

\renewcommand{\theequation}{5.\arabic{equation}}
\setcounter{equation}{0}
\section*{}
\centerline{5. DYNAMICAL FRICTION AND GRAVITATIONAL WAVES}
\vspace{10mm}

In the previous sections
we have shown that in a homogeneous isotropic background
of stars, DF acts so as to bind the binary while progressively
rising $e$.
Eventually, energy and angular momentum losses by gravitational wave
(GW hereafter) emission will determine the terminal evolution of the binary.
These losses depend sensitively on $e$:
For the same energy $E,$ a binary
with $e\sim 0.9$ suffers  a loss a factor $1000$ larger than
for a binary with $e\sim 0$ (Peters \& Mathews 1963).
Due to this dependence,
the transition distance $a_t$ at which the energy loss
by DF equals the power emitted via GW radiation in a binary
is a function of $e$.
$a_t(e)$ can be estimated setting
$\m {\dot E}_{{\rm DF}}\M \simeq \m {\dot E}_{{\rm GW}}\M$ where
\be
\m {\dot E}_{{\rm GW}}\M=\frac{32}{5}\frac{G^4}{c^5}\frac{M_1^5}{a^5}
q^2(1+q)f(e)
\ee
is the energy loss by GWs (Peters \& Mathews 1963); $q\equiv M_2/M_1$
is the mass ratio and the function $f(e)$ reads
\be
f(e) = \frac{1+\frac{73}{24}e^2+\frac{37}{96}e^4}{(1-e^2)^{7/2}}\,.
\ee
Considering equations (5.1) and (3.11),
 the equality in the energy losses leads to
\be
a_t(e)=\left(\frac{16}{5}\frac{G^{5/2}}{c^5\ln\Lambda \,n_o m}\right)^{2/11}
\left[\frac{f(e)}{{\cal K}(e)}\right]^{2/11}\,M^{7/11}.
\ee
Note that $a_t(e)$ depends on $e$ through $f(e)$ and ${\cal {K}}(e)$
and {\it increases\/} with $e.$

The expression of $a_t$ is consistent with
the our treatment of DF only for $a_t(e)\ge \eta r_c$. This inequality
translates into a limit on the parameter $\eta$
as a function of $e,\ q$ and of the core parameters:
\be
\eta<
\left(\frac{16}{5}\frac{G^{5/2}}{c^5\ln\Lambda \,n_o m}\right)^{1/2}
\left[\frac{f(e)}{{\cal K}(e)}\right]^{1/2}\,\left
( {M_c^{7/4}\over r_c^{11/4}}\right )
(1+q)^{7/4}\,.
\ee
$\eta$ depends sensitively on $r_c$ and increases with increasing
$e$. However, for typical parameters of a galaxy core, we find
that $\eta\simeq 10^{-7}$.
In a real scenario DF is effective also at smaller radii and
the actual value of the transition radius depends sensitively
on the dissipative mechanisms for energy and angular momentum transfer
by the ambient stars (see BBR).

\vspace{7mm}
\centerline{5.1 {\it Braking index}}
\vspace{5mm}

The simultaneous presence of the
dissipative forces of DF and GW reaction close to $a_t$
leads,
in this idealized scenario, to the equation for  the evolution of
the semimajor axis $a$
\be
\dot{a}=-\frac{64}{5}\frac{G^3}{c^5}\frac{M^2\mu}{a^3}f(e)
-a {\chi\sigma^2\mu^{3/2}\over 2^{1/2}\pi }{ {\cal K}(e)\over E^{3/2}}
\ee
derived using equation (5.1), and (3.11) for $x\gg 1.$

This evolution equation can formally be written as
\be
\dot{a}=-\frac{\alpha}{a^3}\left(1+\frac{\gamma}{\alpha} a^{11/2}\right)\,,
\label{agwdf}
\ee
where
\be
\alpha  =  \frac{64}{5}\frac{G^3}{c^5}\,M^2 \mu f(e)
\ee
and
\be
\gamma={2\chi\sigma^2\over \pi (GM)^{3/2} }{\cal K}(e).
\ee

With use of Kepler's third law, we can relate the evolution of $a(t)$
to the  frequency $f(t)=1/P$ of orbital motion of the
system:
\be
f=\frac{1}{2\pi}\left(\frac{GM}{a^3}\right)^{1/2}\,.
\label{kepl}
\ee
Due to orbit decay, we further
have from  (\ref{agwdf}) and (\ref{kepl})
\be
\dot{f}=
\frac{3}{2\pi}\sqrt{GM}\left(\frac{\alpha}{a^{11/2}}+\gamma \right)\,,
\label{fdot}
\ee
\be
\ddot{f}=
\frac{33}{4\pi}\sqrt{GM}\frac{1}{a^{13/2}}
\left(\frac{\alpha}{a^3}+\gamma a^{5/2}\right).
\label{fdot2}
\ee
We can thus introduce the  braking index
$n$
\be
n=\frac{f\ddot{f}}{\dot{f}^2}
\ee
that takes the following form
\ba
n & = & \frac{11}{3}\left[1+\frac{\gamma}{\alpha}
a^{11/2}\right]^{-1}\nonumber\\
& = & n_{{\rm GW}}\left[1+\left(\frac{a}{a_t(e)}\right)^{11/2}\right]^{-1}\,,
\ea
where
$n_{{\rm GW}}={11}/{3}$ is the value of the index in absence of DF.

In principle, a measurable deviation from 11/3
would provide an indirect way for testing the ambient star medium.
In Figure 6 the braking index $n$ (eq. [5.13]) is shown
as a function of  the BH
binary separation for various $e$. It appears that
a large deviation from 11/3 occurs close to $a_t$, where the
two forces are competing.
Above $a_t$ the braking index is quite small since DF
is the cause of the deceleration of the orbit. Below
$a_t$ on the contrary the index is close to $n_{{\rm GW}}$, since GW
losses dominate.

Recently, Giampieri (1993) considered the deviation from the value
$n_{{\rm GW}}$ induced by mass transfer.
The correction term in that case has a sign
that depends on the mass difference of the two BHs and on the details
of mass transfer.
The calculation of $n$
is here proposed as example for suggesting a way to probe
the environment of a BH binary.
The validity of such approach depends however on the relevance of the
process considered in the region where
an outgoing GW signal is detectable.
For the case
of DF,  the GW signal about $a_t$
would be still very weak and of extremely low-frequency.
The possibility of revealing
GWs about $a_t$ is thus quite remote. On the contrary, mass transfer processes
occurring close to the gravitational
radius of the BHs in the binary, could appear in the measure of $n$ as
deviation from the pure gravitational effect.

At low enough BH separations
the  energy loss to the field stars  via single
encounters lasting less than a period $P$, might produce a
 sudden rise in the GW frequency $f$, with a
$\Delta f/f\sim m/M$
for an impact parameter $\sim a$.
For smaller impact parameters, of the order of the gravitational
radius $R_G(M_2)=
2GM_2/c^2$, the `` glitch" might have a magnitude
$\Delta f/f$ as large as $\sim (m/M)a/R_G(M_2).$
The potential observability of these phenomena would be of considerable
importance.

\section*{}
\centerline{6. CONCLUSIONS}
\vspace{10mm}

We have explored the long term evolution of a
black hole pair under the
action of dynamical friction by a stationary
homogeneous and isotropic background
of light particles. The result applies to the case
of a BH pair in the center of a galactic nucleus,
in regions where the stellar distribution can be approximated
as homogeneous.

Future work
should focus on the
study of DF in a nonuniform anisotropic star background
with the aim at exploring the subsequent binary evolution (BBR; PR).

\vspace{10mm}
\noindent
We gratefully thank B. Bertotti for enlightening discussions and constant
encouragement.
This research was carried out with the financial support from the
Italian Ministero dell'Universit\`a e della Ricerca Scientifica e
Tecnologica.

\newpage

\appendix
\renewcommand{\theequation}{A.\arabic{equation}}
\setcounter{equation}{0}
\section*{}
\centerline{APPENDIX A}
\vspace{10mm}

In this Section we outline the analytical method
used to calculate $\m\dot{S}\M$ in the case of the
Keplerian potential. We proceed along three steps:

(1) First we specify the instantaneous motion of the
secondary BH, described by the velocity vector $\gras {v}$
and the separation vector $\gras {r}$.  Since DF
is a weak perturbing force, the motion is determined by the driving
acceleration force of the BH primary (see $\S \,\, 2$).
$\gras {v}$
and $\gras {r}$
are expressed in term of the orbital phase $\psi$: They are
defined uniquely by the instantaneous values of $E,J,a$ and $e$.
For $\ r\equiv {\gras {r}}/{\hat {\gras {r}}}$ we have
\be
{ r}=\frac{a(1-e^2)}{1+e\cos\psi}\,.
\label{raggio}
\ee
The velocity is decomposed along the radial and tangential directions
to yield the following expressions:
\ba
v_r & = & \gras{v}\cdot\hat{\gras{r}}=\dot{r}\nonumber\\
& = & \frac{GM\mu}{J}e\sin\psi=\left[\frac{GM}{a(1-e^2)}\right]^{1/2}
e\sin\psi\,,
\ea
\ba
v_t & = & \frac{J}{\mu r}\nonumber\\
& = & \frac{GM\mu}{J}(1+e\cos\psi)=\left[\frac{GM}{a(1-e^2)}\right]^{1/2}
(1+e\cos\psi)\,.
\ea
Furthermore,
\be
v^2 = \frac{GM}{a(1-e^2)}(1+2 e \cos\psi+e^2)\,.
\ee

\par
(2) Secondly, the perturbing force
${\dot{\gras{v}}}_{{\rm DF}}$
is  decomposed along three orthonormal directions
defined by the BHs separation vector $\gras {r}$, the velocity $\gras{v}$
and $\gras {J}$.
These components are indicated with
$R$ acting along $\gras{r}$, $W$
along $\gras {J},$  and $T$ in the orbital plane.
The three components of the frictional deceleration
given by equation (2.1) are accordingly to Bertotti \& Farinella (1990)
(BF hereafter)
\ba
R & = & \dot{ \gras{v}}_{{\rm DF}} \cdot \frac{\gras{r}}{r}\nonumber\\
& = & -\chi g(x) \frac{v_r}{v}\nonumber\\
& = & -\chi g(x) \frac{e\sin\psi}{(1+2e\cos\psi+e^2)^{1/2}}\,,
\ea
\ba
T & = & \dot {\gras{v}}_{{\rm DF}}
\cdot \frac{\gras{J}\times\gras{r}}{Jr}\nonumber\\
& = & -\frac{\mu}{J} \chi g(x)
\frac{v_t^2 r}{v}\nonumber\\
& = & -\chi g(x) \frac{1+e\cos\psi}{(1+2e\cos\psi+e^2)^{1/2}}\,,
\ea
\ba
W &  = & \dot {\gras{v}}_{{\rm DF}} \cdot \frac{\gras{J}}{J}\nonumber\\
& = & 0\,.
\ea

\vspace{7 mm}
In particular,
for $x\ll 1$ we have :
\ba
R & = & -\frac{4\chi}{3(2\pi)^{1/2}\sigma}
\left[\frac{GM}{a(1-e^2)}\right]^{1/2} e \sin \psi \nonumber\\
& & +
\frac{2\chi}{5(2\pi)^{1/2}\sigma^3}
\left[\frac{GM}{a(1-e^2)}\right]^{3/2}e\sin\psi (1+2 e \cos\psi+e^2)\nonumber\\
& & + o(x^3)\,,
\ea
\ba
T & = & -\frac{4\chi}{3\sqrt{2\pi}\sigma}
\left[\frac{GM}{a(1-e^2)}\right]^{1/2}(1+e\cos\psi)
 \nonumber\\
& & + \frac{2\chi}{5(2\pi)^{1/2}\sigma^3}
\left[\frac{GM}{a(1-e^2)}\right]^{3/2}
(1+e\cos\psi) (1+2 e \cos\psi+e^2)\nonumber\\
& & + o(x^3)\,,
\ea

\vspace{7 mm}

and for $x\gg 1$:
\be
R=-2 \chi\sigma^2\frac{a(1-e^2)}{GM}\frac{e\sin\psi}
{(1+2 e \cos\psi+e^2)^{3/2}}\,,
\ee
\be
T=-2 \chi\sigma^2\frac{a(1-e^2)}{GM}\frac{1+e\cos\psi}
{(1+2 e \cos\psi+e^2)^{3/2}}\,.
\ee

(3) As third step, we express $\dot E$, $\dot J$ and $\dot e$
in terms of $R$ and $T.$
Following closely BF we have
\ba
\dot{E} & = & \mu \dot{\gras{v}}_{{\rm DF}} \cdot \gras{v}
\nonumber\\
& = & {GM\over J}\left [ Re\sin \psi +T(1+e\cos\psi)\right]\,,
\label{enper}
\ea
\ba
\dot{J} & = &
\mu ( {\gras {r}}\times \dot {\gras{v}}_{{\rm DF}} )
\cdot { {\gras {J}}\over { J } }
\nonumber\\
& = & \mu rT\,,
\label{maper}
\ea
and
\be
\dot{e}=\frac{P(1-e^2)^{1/2}}{2\pi a}
\left [R\sin\psi+T\,\frac{e\cos^2\psi+2\cos\psi+e}{1+e\cos\psi}\right ]
\label{eper}\,,
\ee
with $P$ the Keplerian period.
The derivatives written above are functions of the orbital phase $\psi$.
In Appendix B we carry out their mean, over the orbital motion.

\newpage
\setcounter{equation}{0}
\renewcommand{\theequation}{B.\arabic{equation}}
\section*{}
\centerline{APPENDIX B}
\vspace{10mm}

In this Appendix we sketch the derivation of the
equations for the long term evolution of
$E$ and $J$. The calculations involve the explicit
integration of equations (A.12) and (A.13) over the orbital phase $\psi$.
Below we indicate the main steps.

Along a Keplerian orbit, the phase $\psi$ is related to the
time coordinate by the relation
\be
dt = \frac{P}{2\pi}(1-e^2)^{3/2}\frac{1}{(1+e\cos\psi)^2}d\psi\,.
\ee
This equation is used to transform  the time  integral (3.3)
to an integral over $\psi$.
The mean variation of energy $E$ thus reads:
\be
\langle \dot{E} \rangle = -\frac{\mu\chi}{2\pi}
\left(\frac{GM}{a}\right)^{1/2}(1-e^2)\int_0^{2\pi}g[x(\psi)]
\frac{(1+2 e\cos\psi+e^2)^{1/2}}{(1+e\cos\psi)^2}d\psi\,;
\label{aenergy}
\ee
equivalently for $J$:
\be
\langle \dot{J} \rangle = -\frac{\mu\chi}{2\pi}a
(1-e^2)^{5/2}\int_0^{2\pi}g[x(\psi)]
\frac{1}{(1+2 e\cos\psi+e^2)^{1/2}(1+e\cos\psi)^2}d\psi\,.
\label{aangmom}
\ee
In the two separate
limits $x\ll 1 $ and $x\gg 1,$ the above integrals can be evaluated
analytically.  For arbitrary $x(\psi)$, the calculation is carried
out numerically
and Figure 4 summarizes the results of the integration.
The analytical expressions in the two asymptotic limits
were used as a test on numerical accuracy.

\vspace{5mm}
\centerline{(i) $x\ll 1$}
\vspace{5 mm}

In the calculation that follows, the frictional drag $g(x)$
is given by equation (2.5). According to the expression of $g(x),$
we distinguish in $\m \dot E \M$
the contribution associated to the first
leading term $(\propto x$) in the expansion of $g(x)$
(indicated with label 1) and the subsequent term ($\propto x^3$)
(indicated with label 3).
The frictional drag $g(x)$ is known function of $\psi$ through $x$
given by equations (2.3) and (A.4). The resulting integral will formally read:
\be
\langle \dot{E} \rangle = \langle \dot{E} \rangle_1 +
\langle \dot{E} \rangle_3 + o(x^3)\,.
\ee
As for the energy, the variation of $J$ splits in two terms
that will be evaluated separately
\be
\langle \dot{J} \rangle = \langle \dot{J} \rangle_1 +
\langle \dot{J} \rangle_3 + o(x^3)\,.
\ee
{}From equations (\ref{aenergy}), (\ref{aangmom}) and (2.5) we find:
\be
\langle \dot{E} \rangle_1 = \frac{2\chi}{3\pi(2\pi)^{1/2}\sigma}\frac{GM\mu}{a}
(1-e^2)^{1/2}I_1(e)\,,
\label{aenergy1}
\ee
\be
\langle \dot{E} \rangle_3 = \frac{\chi\mu}{5\pi(2\pi)^{1/2}\sigma^3}
\left(\frac{GM}{a}\right)^2\frac{I_{2}(e)}{(1-e^2)^{1/2}}\,,
\label{aenergy3}
\ee
\be
\langle \dot{J} \rangle_1 =-\frac{2\chi\mu}{3\pi(2\pi)^{1/2}\sigma}
(GM a)^{1/2}(1-e^2)^2 I_0(e)\,,
\label{amoman1}
\ee
\be
\langle \dot{J} \rangle_3 = \frac{\chi\mu}{5\pi(2\pi)^{1/2}\sigma^3}
\left(\frac{G^3 M^3}{a}\right)^{1/2}(1-e^2)I_{1}(e)\,,
\label{amoman3}
\ee
where
\be
I_n(e) = \int_0^{2\pi}
\frac{(1+2 e\cos\psi+e^2)^n}{(1+e\cos\psi)^2}d\psi\,,
\ee
After some straightforward algebra we have:
\be
I_0(e) = \frac{2\pi}{(1-e^2)^{3/2}}\,,
\ee
\be
I_1(e) = \frac{2\pi}{(1-e^2)^{1/2}}\,,
\ee
\be
I_2(e) = 2\pi\left[4-3(1-e^2)^{1/2}\right]\,,
\ee
Substituting the values of $I_0$, $I_1$ and $I_2$ in (\ref{aenergy1}),
(\ref{aenergy3}), (\ref{amoman1}) and (\ref{amoman3}) we obtain
equation (3.8) for $\m \dot{E} \M_1$, equation (3.9) for
$\m \dot{J} \M_1$ and
\be
\langle \dot{E} \rangle_3 = \frac{2\chi\mu}{5(2\pi)^{1/2}\sigma^3}
\left(\frac{GM}{a}\right)^2\frac{4-3(1-e^2)^{1/2}}{(1-e^2)^{1/2}}\,,
\ee
\be
\langle \dot{J} \rangle_3 = \frac{2\chi\mu}{5(2\pi)^{1/2}\sigma^3}
\left(\frac{G^3 M^3}{a}\right)^{1/2}(1-e^2)^{1/2}\,
\ee
respectively.

\vspace{5mm}
\centerline{(ii) $x\gg 1$}
\vspace{5 mm}

In the limit $x\gg 1,$ equation  (\ref{aenergy})
becomes:
\be
\m \dot{E} \M = -\frac{\mu\chi\sigma^2}{\pi}
\left(\frac{a}{GM}\right)^{1/2}{\cal K}(e)\,,
\ee
where
\be
{\cal K}(e) = (1-e^2)^2\frac{1}{2\pi}I_{-1/2}(e)\,.
\label{intk}
\ee

We can perform the integration analytically, expanding (\ref{intk}) about
 $e\sim 0$; it is found
\be
{\cal K}(e) = \left[1+\frac{3}{4}e^2+\frac{21}{64}e^4+o(e^4)\right]\,.
\label{intka}
\ee
The behaviour of ${\cal K}(e)$ is given in Figure B1.

Analogously for $\m \dot{J} \M$, from equation (\ref{aangmom}) we find:
\be
\m \dot{J} \M = -\frac{\mu\chi\sigma^2}{\pi}
\frac{a^2}{GM}{\cal Z}(e)\,,
\ee
where
\be
{\cal Z}(e) = (1-e^2)^{7/2}\frac{1}{2\pi}I_{-3/2}(e)\,.
\label{intz}
\ee
Expanding equation (\ref{intz}) for $e\ll 1$ we have:
\be
{\cal Z}(e) = \left[1+\frac{13}{4}e^2+\frac{157}{64}e^4+o(e^4)\right]\,.
\label{intza}
\ee
The behaviour of ${\cal Z}(e)$ is given in Figure B2.

\setcounter{equation}{0}
\renewcommand{\theequation}{C.\arabic{equation}}
\section*{}
\centerline{APPENDIX C}
\vspace{10mm}

Now we derive the formula of the growth of the eccentricity. Substituting
in equation (A.14) the equations (A.5) and (A.6),  and subsequently
averaging we
find:
\ba
\langle \dot{e} \rangle & = & -\frac{\chi}{\pi}(1-e^2)^2
\left(\frac{a}{GM}\right)^{1/2}\nonumber\\
& & \int_0^{2\pi}g[x(\psi)]
\frac{e+\cos\psi}{(1+2 e\cos\psi+e^2)^{1/2}(1+e\cos\psi)^2}d\psi\,.
\label{aecc}
\ea

\vspace{5mm}
\centerline{(i) $x\ll 1$}
\vspace{5 mm}

Retaining, as in the case of $E$ and $J$,  terms
to order $x^3$ in $g[x(\psi)],$
we write $\langle \dot{e} \rangle$ in the form:
\be
\langle \dot{e} \rangle = \langle \dot{e} \rangle_1 +
\langle \dot{e} \rangle_3 + o(x^3)\,.
\ee
Using the asymptotical expression of $g(x)$, we obtain:
\be
\langle \dot{e} \rangle_1 = -\frac{4\chi}{3\pi(2\pi)^{1/2}\sigma}
(1-e^2)^{3/2} Q_0(e)\,,
\ee
\be
\langle \dot{e} \rangle_3 = \frac{2\chi}{5\pi\sigma^3(2\pi)^{1/2}}
\left(\frac{GM}{a}\right)(1-e^2)^{1/2}Q_1(e)\,,
\ee
where
\be
Q_n(e) = \int_0^{2\pi}
\frac{(e+\cos\psi)(1+2 e\cos\psi+e^2)^n}{(1+e\cos\psi)^2}d\psi\,.
\ee
After some straightforward algebra we obtain:
\be
Q_0(e) = 0
\ee
and
\ba
Q_1(e)
= \frac{4\pi}{e}\left[1-(1-e^2)^{1/2}\right]\,.
\ea

\vspace{5mm}
\centerline{(ii) $x\gg 1$}
\vspace{5mm}

Substituting equation (2.6) into (\ref{aecc}) we obtain:
\be
\m \dot{e} \M = 2\frac{\chi\sigma^2}{\pi}
\left(\frac{a}{GM}\right)^{3/2}{\cal W}(e)\,,
\ee
where
\be
{\cal W}(e) = - (1-e^2)^3 \frac{1}{2\pi}Q_{-3/2}(e)\,.
\ee
Expanding ${\cal W}(e)$ for $e\ll 1$ we find:
\be
{\cal W}(e) = \frac{3}{2} e\left[1+\frac{3}{8}e^2+o(e^2)\right]\,.
\ee
The behaviour of ${\cal W}(e)$ is given in Figure C1.


\newpage
\section*{}
\centerline{REFERENCES}
\vspace{10mm}
\begin{description}
\item Barnes, J.E., \& Hernquist, L. 1992,
ARA\&A, 30, 705
\item Begelman, M.C., Blandford, R.D., \& Rees, M.J. 1980, Nature, 287,
307 (BBR)
\item Bekenstein, H., \& Zamir, R. 1990, ApJ, 359, 427
\item Bertotti, B., \& Farinella, P. 1990, Physics of the Earth and the
Solar System, Kluwer Academic Publishers (BF)
\item Binney, J., \& Tremaine, S. 1987, Galactic Dynamics,
Princeton University Press
\item Bland-Hawthorn, J., Wilson, A.S., \& Tully R.B. 1991, ApJ, 371, L19
\item Chandrasekhar, S. 1943, ApJ, 97, 255
\item Danzmann, K., et al. 1993, MPQ 177, Proposal for a Laser-Interferometric
Gravitational Wave Detector in Space
\item Dressler, A. 1989, in IAU Stmp. 139, Active Galactic Nuclei, ed. D.E.
Osterbrock, \& J.S. Miller (Dordrecht: Kluwer), 217
\item Estabrook, F.B., \& Wahlquist, H.D. 1975, Gen. Rel. Grav., 6, 439
\item Farouki, R. T., \& Salpeter, E. E. 1982, ApJ, 253, 512
\item Frank, J., \& Rees, M.J. 1976, MNRAS, 176, 633
\item Fukushige, T., Ebisuzaki, T., \& Makino, J. 1992a,
PASJ, 44, 281
\item Fukushige, T., Ebisuzaki, T., \& Makino, J. 1992b, ApJ, 396, L61
\item Gaskell, C.M. 1985, Nature, 315, 386
\item Giampieri, G. 1993, preprint
\item Gould, A. 1993, ApJ, 379, 280
\item Governato, F., Colpi, M., \& Maraschi, L. 1993, in preparation
\item Hellings, R.W., et al. 1993, JPL Engineering Memorandum 314-569
\item Hills, J.G. 1983, AJ, 88, 1269
\item Makino, J., Fukushige, T., Okumura, S.K., \& Ebisuzaki, T.
1993, PASJ, 45, 303
\item Maoz, E. 1993, MNRAS, 263, 75
\item Mikkola, S., \& Valtonen, M.J. 1992, MNRAS, 259, 115
\item Peters,P.C. 1964, Phys. Rev., 136, B1244
\item Peters, P.C., \& Mathews, J. 1963, Phys. Rev., 131, 435
\item Polnarev, A.G., \& Rees,M.J. 1994, A\&A, 283, 301
\item Rees, M.J., 1990, Science, 247, 817
\item Roos, N. 1981, A\&A, 104, 218
\item Roos, N. 1988, ApJ, 334, 95
\item Roos, N., Kaastra, J.S., \& Hummel, C.A. 1993, ApJ, 409, 130
\item Smith, H. 1992, ApJ, 398, 519
\item Thorne, K.S., \& Braginsky, V. B. 1976, ApJ, 204, L1
\item Thorne, K.S. 1992, in Recent Advances in General Relativity
A. Janis and J. Porter, (Birkhauser, Boston)
\item Valtaoja, L., Valtonen, M.J., \& Byrd, G.G. 1989, ApJ, 343, 47
\end{description}

\vspace{40mm}

\section*{}
\centerline{FIGURE CAPTIONS}
\vspace{10mm}
\parindent=0pt
\smallskip
\noindent
{\bf Figure 1}. Time scale $\tau_>$
(in units of $10^8$ yrs) describing the orbital decay of the secondary
BH in the Keplerian potential of the primary,
as a function of $e$ for
$x\gg 1$. Top curve denotes the time scale for energy loss
$\tau_{>}^E$ (eq. [3.14]); bottom curve
denotes the time scale for angular momentum loss
$\tau_{>}^J$ (eq. ([3.15]).

\smallskip
\noindent
{\bf Figure 2}. Time scale for the growth
of the eccentricity $e$, $\tau_{<}^e$ (in units of $10^8$yrs)
as a function of $e$, for $x\ll 1.$ The secondary
BH is in the Keplerian potential of $M_1.$

\smallskip
\noindent
{\bf Figure 3}. Time scale for the growth
of the eccentricity $e$, $\tau_{>}^e$ (in units of $10^8$yrs)
as a function of $e$, for $x\gg 1.$ The secondary
BH is in the Keplerian potential of $M_1.$

\smallskip
\noindent
{\bf Figure 4}. $\m {\dot  E}\M/\m {\dot J}\M$
in units of $-2E/J$ as a function of $e$
for selected values of $x_{\rm {cir}}$ ($0.1$, $0.5$, $1$, $10$)
summarizing the results of
the eccentricity increase for the orbital evolution
of the secondary BH in the Keplerian potential of $M_1.$
Dots indicate the analytical expression
of $\m {\dot  E}\M/\m {\dot J}\M$ as given by eq. [3.25] (top curve)
and eq. [3.29] (bottom curve).

\smallskip
\noindent
{\bf Figure 5}. The eccentricity evolution
for paths with initial semimajor axis $a_i=20$ pc. The curves give
$e$ as a function of $a$ (in pc)
for initial values of the eccentricity $e_i=0.05, 0.03, 0.01$
starting from the top respectively. The dashed lines indicate evolutionary
paths with pericenter distance $r_{{\rm P}}<r_{{\rm cusp}}$.

\smallskip
\noindent
{\bf Figure 6}. Braking index $n$ as a function of the semimajor
axis $a$ in unit of $a_t(e=0),$ as
defined by eq. (5.13)
for selected values of $e=0, 0.4,0.7$ starting from left, respectively.

\smallskip
\noindent
{\bf Figure B1}. ${\cal K}$
versus $e$, as defined by eq. (B.17): dots denote the
analytical expansion (eq. [B.18]).

\smallskip
\noindent
{\bf Figure B2}. ${\cal Z}$ versus $e$, as defined by eq. (B.20):
dots denote the analytical expansion (eq. [B.21]).

\smallskip
\noindent
{\bf Figure C1}. ${\cal W}$ versus $e$, as defined by eq. (C.9):
dots denote the analytical expansion (eq. [C.10]).

\end{document}